\documentclass[aps,pra,reprint,superscriptaddress,floatfix]{revtex4-2}
\usepackage{soul}
\usepackage{amsmath,amssymb,cases,bm} 
\usepackage{graphicx}
\graphicspath{{figures/}}

\usepackage[colorlinks=true,urlcolor=blue,breaklinks=true]{hyperref}
\usepackage[dvipsnames]{xcolor}

\begin{document}

\title{Bloch-sphere rotations in driven double-well for ultracold atoms}

\author{Michele Modugno}
\affiliation{\mbox{Department of Physics, University of the Basque Country (EHU), 48080 Bilbao, Spain }}
\affiliation{\mbox{IKERBASQUE, Basque Foundation for Science, 48009 Bilbao, Spain}}
\affiliation{\mbox{EHU Quantum Center, University of the Basque Country (EHU), 48940 Leioa, Spain}}

\author{Marco Fattori}
\affiliation{\mbox{Dipartimento di Fisica e Astronomia, Universit\`a di Firenze, 50019 Sesto Fiorentino, Italy}}
\affiliation{\mbox{Istituto Nazionale di Ottica, Consiglio Nazionale delle Ricerche (CNR-INO), Largo Enrico Fermi 6, 50125 Firenze, Italy}}
\affiliation{\mbox{European Laboratory for Nonlinear Spectroscopy (LENS), Via N. Carrara 1, 50019 Sesto Fiorentino, Italy}}

\author{Giulio Pettini}
\affiliation{\mbox{Dipartimento di Fisica e Astronomia, Universit\`a di Firenze, 50019 Sesto Fiorentino, Italy}}
\affiliation{\mbox{INFN Sezione di Firenze, Italy}}

\begin{abstract}
We show that, by using suitable protocols for a non-interacting condensate in a driven double-well, one can achieve controlled rotations about arbitrary axes in the equatorial plane of the Bloch sphere, composed with fast rotations about the $z$-axis.
Specifically, we investigate the dynamics induced by a spatially linear time-periodic potential, by means of numerical simulations.
We also provide an explicit two-level model that accurately captures the microscopic evolution of the driven system, with the full time-dependent evolution operator obtained using a Floquet-based approach.
The analysis is carried out using as a reference a recently realized experimental platform consisting of arrays of double-well potentials based on Beat-Note Superlattices, to which the proposed control scheme is directly applicable.
\end{abstract}

\maketitle

\section{Introduction}

The coherent manipulation of quantum states lies at the heart of atom interferometry and quantum-enhanced sensing. Recent advances in trapped Bose–Einstein condensate (BEC) interferometers have demonstrated that non-interacting condensates confined in double-well potentials can realize high-contrast beam splitters, long coherence times, and full Mach–Zehnder interferometric sequences \cite{petrucciani2026}. In such systems, the two spatial modes of the double well naturally define an effective two-level system, where coherent tunneling between the wells provides controlled rotations around the $x$-axis of the Bloch sphere. These results establish trapped double-well interferometers as a promising platform for precision sensing and quantum technologies, while simultaneously raising the question of how to achieve full coherent control of the associated pseudospin dynamics.

While rotations around the $x$-axis are naturally generated by tunneling, many advanced protocols for quantum sensing and quantum control require access to a second orthogonal axis. Rotations around the $y$-axis are a key ingredient for the implementation of composite pulse sequences, dynamical decoupling schemes, robust beam splitters, and error-compensation protocols designed to suppress imperfections in the interferometric operations or uncontrolled energy biases between the two wells \cite{Cummins}. In conventional two-level systems driven by electromagnetic fields, arbitrary rotation axes can be obtained by simply controlling the phase of the driving field. For matter-wave interferometers based on spatially separated modes, however, the realization of analogous operations is considerably less straightforward.

The possibility of engineering effective tunneling amplitudes through time-periodic potentials has been extensively investigated in optical lattices and represents one of the most successful applications of Floquet engineering. Periodic modulation techniques have enabled coherent destruction of tunneling \cite{Grossmann1991}, photon-assisted tunneling \cite{Sias2008} and artificial gauge fields \cite{Aidelsburger2011,Goldman2014}.
Despite these remarkable developments, the implementation of controlled rotations around the $y$-axis in a double-well interferometer has received comparatively little attention. In particular, while Floquet techniques have been successfully employed to modify the magnitude and sign of the tunneling amplitude, their direct connection with the geometric control of the Bloch sphere dynamics in a double-well system remains largely unexplored.

Beyond its practical relevance, the double-well potential constitutes one of the simplest quantum systems in which coherent matter-wave dynamics can be studied. It also offers a transparent setting in which the fundamental concepts of Floquet engineering can be directly connected to experimentally relevant interferometric operations.

A recent experiment \cite{impertro2024} has demonstrated programmable rotations in optical double wells with single-site resolution, using tunnel coupling and a potential energy bias as rotations about the $x$- and $z$-axes of the Bloch sphere. 
This approach was applied to the measurement of kinetic operators in optical lattices. 
The protocol relies on decoupled-well $z$-rotations, which are inherently more sensitive to potential fluctuations than $x$-rotations. 
The present work is complementary to this approach, in that we demonstrate how time-periodic driving of double-well systems with suitable stop-and-go protocols allows for controlled rotations about arbitrary axes in the $xy$-plane of the Bloch sphere, composed with fast rotations about the $z$-axis. 
In particular, we consider rotations about the $y$-axis, which we analyze through numerical simulations. 
We focus on the regime of negligible interparticle interactions and appropriate driving frequencies, where the dynamics reduce to an effective two-level description. 
This enables an analytical Floquet approach that provides the full time-dependent evolution operator and captures the microscopic dynamics of the driven system within a simple and accurate framework.

The paper is organized as follows. In Sect.~\ref{sec:system} we introduce the double-well system realized by a Beat-Note Superlattice and its corresponding two-level description. In Sect.~\ref{sec:drivendw} we present the two driving protocols considered in this work and discuss the resulting dynamics, obtained from numerical simulations. These results are then interpreted within the two-level formalism introduced in Sect.~\ref{sec:tlm}, where we compute the full time-dependent evolution operator using the Floquet formalism within the Magnus approximation, which accurately reproduces the numerical findings. 
Finally, conclusions are drawn in Sect.~\ref{sec:concl}. In the Appendix, we provide additional details of the derivations.

\section{Double-well system}
\label{sec:system}

In this work, we consider a non-interacting condensate in a double-well potential. 
As an illustrative example, we focus on a specific implementation with a non interacting BEC of $^{39}$K atoms trapped in Beat-Note Superlattices (BNSLs), as realized in recent experiments \cite{masi2021bnsl,masi2021}. 
By superimposing two or more optical lattices (OLs) with suitably chosen wavelengths $\lambda_i$, BNSLs offer the possibility of creating optical superlattices with arbitrarily large spacing. 
In particular, by combining three OLs fulfilling the conditions $n\lambda_2=(n+1)\lambda_1$ and $(n-1)\lambda_3=(n+1)\lambda_1$, with $n$ an integer, one can realize an optical potential for atoms that behaves as an effective array of double wells \cite{petrucciani2025}.

\begin{figure}[t]
\centerline{\includegraphics[width=0.95\columnwidth]{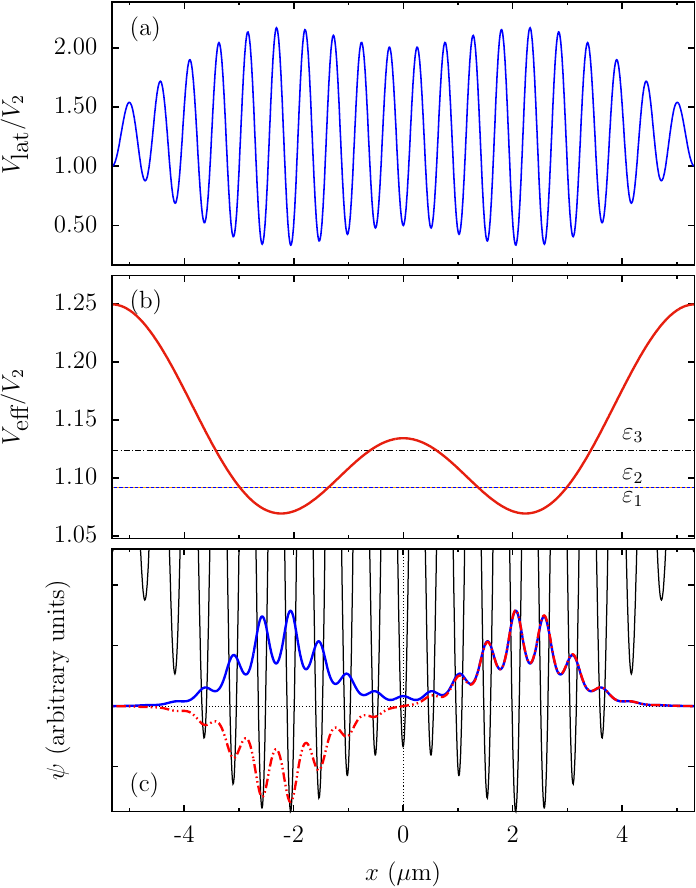}}
\caption{(a) Plot of the BNSL potential in Eq.~\eqref{eq:beatnotepot}.
(b) Effective approximation \eqref{eq:effpot}, corresponding to a symmetric double well with a spatial separation between the two minima $a \simeq 4.436~\mu$m.
The first three energy levels $\varepsilon_i$ ($i=1,2,3$) of $H_0$ in Eq.~\eqref{ham} are also shown, displayed as horizontal lines.
(c) Spatially symmetric (blue solid line) and antisymmetric (red dot-dashed line) low-lying eigenfunctions. The potential (black solid line) is included for reference.}
\label{fig1}
\end{figure}

With this in mind, we start from the Hamiltonian
\begin{equation}
H_0(x) = -\displaystyle{\frac{\hbar^2}{2m}\frac{d^{2}}{dx^{2}}} + V_{B3}(x),
\label{ham}
\end{equation}
with
\begin{equation}
	V_{B3}(x) = \sum_{i=1,2} V_{i}\sin^{2}(q_{i} x) + V_{3}\sin^{2}(q_{3} x + \pi/2),
\label{eq:beatnotepot}
\end{equation}
where $q_i=2\pi/\lambda_i$ ($i=1,2,3$). 
Here we consider a lattice with a parity-symmetric unit cell, as shown in Fig.~\ref{fig1}(a). We prepare the system in a single cell and work in a regime of parameters where tunneling to neighboring cells is negligible, so that the dynamics is effectively confined to a single double-well 
\footnote{We have explicitly checked by numerical simulations that tunneling between neighboring cells is negligible over the time scales considered in this work.}.
At low lattice depths, the overall confinement is well approximated by the effective double-well potential in Fig.~\ref{fig1}(b), given by \cite{petrucciani2025}
\begin{equation}
V_{\mathrm{eff}}(x) = 
\frac{V_2(V_3 - V_1)}{16E_{R_{2}}}\cos({\bar{q}} x) 
+ \frac{V_1V_3}{16E_{R_{2}}}\cos(2{\bar{q}} x),
\label{eq:effpot}
\end{equation}
modulo a constant term, where ${\bar{q}}=q_1-q_3$ and $E_{R_{2}}=\hbar^2q_2^2/(2m)$.

To provide a specific example, we choose $\lambda_2 = 1064.5$~nm, $n=20$, $V_1 = V_2 = 400$~nK, and $V_3 = V_2/2$, as in Refs. \cite{masi2021bnsl,masi2021,petrucciani2025}. With this choice of parameters, the spectrum of the resulting double-well consists of two quasi-degenerate low-lying levels well separated from the excited ones, also shown in Fig.~\ref{fig1}(b).
Specifically, defining $\nu_{ij} = \Delta E_{ij}/h$, we have $\nu_{12}=1.695$~Hz, whereas the gap to the third energy level is $\nu_{23}=265.5$~Hz. 
The eigenfunctions $\psi_{i}$ ($i=1,2$) of the first two levels are shown in Fig.~\ref{fig1}(c).

We remark that the results presented in the following are not restricted to this specific parameter set and apply more generally within the same symmetry class, under the driving regime discussed below.

\subsection{Two-level dynamics}
\label{sec:tldynamics-free}

For the discussion in the next section, it is convenient to first review the main features of the static model described by $H_0$, focusing on its two lowest levels. As usual, we consider the normalized combinations of the two low-lying eigenfunctions,
\begin{equation}
\psi_{R,L}=\displaystyle{\frac{\psi_1\pm\psi_2}{\sqrt{2}}},
\end{equation}
which are localized in the right ($R$) and left ($L$) sub-wells and can be associated with the $\sigma_z$ eigenstates of the two-level Hamiltonian
\begin{equation}
H^{\mathrm{tl}}_{0} = -J \sigma_{x},
\label{hamsigmax}
\end{equation}
with a (real) bare tunneling amplitude $J=\Delta E_{12}/2$.

Initializing the system in either $\psi_R$ or $\psi_L$, the dynamics consists of oscillations between the two sub-wells with Rabi period $T_R=h/(2J)=1/\nu_{12}$, which are experimentally observable. 
The analytic formula predicts $T_{R} = 590$ ms, in excellent agreement with the behavior of the full Hamiltonian \eqref{ham}, as confirmed by numerical simulations \footnote{The observed discrepancy between full numerical results and the prediction of the analytic formula is less than $1\%$, for $V_{2}$ ranging from $100$ to $400$~nK.}.
Representing the state of the system by the unit vector on the Bloch sphere, 
\begin{equation}
    \bm{n}(t)=\langle\psi(t)|{\bm{\sigma}}|\psi(t)\rangle,
\end{equation}
with $\bm{\sigma}=(\sigma_{x}, \sigma_{y}, \sigma_{z})$, the dynamics corresponds to a rotation around the $x$-axis. In particular, after a time $t=T_R/4$, the system evolves into an eigenstate of $\sigma_y$, manifesting as an equiprobable density distribution between the $L$ and $R$ sub-wells. This observation, although simple, will be relevant for the following discussion in the presence of a time-periodic drive.

\section{Driven double-well}
\label{sec:drivendw}

In this section we discuss how to control the population imbalance among the two sub-wells through the realization of a time-periodic Hamiltonian of the form
\begin{equation}
H(x,t) = H_0(x) + V(x,t),
\label{hamtimeperiodic}
\end{equation}
by means of appropriate stop-and-go protocols with the potential
\begin{equation}
V(x,t) = -F_{0} x\sin(\omega t),
\label{eq:pott}
\end{equation}
which is switched on at $t=0$. The system is assumed to be undriven for negative times. 
The dynamics considered here is obtained by numerically integrating the time-dependent Schrödinger equation for the Hamiltonian $H(t)$.

It is worth remarking that an oscillating potential of this form is generally considered in the literature to be realized using driven lattices, namely by periodically shifting the position of the lattice and transforming to the comoving reference frame \cite{graham1992,drese1997,madison1998,eckardt2005,struck2012,eckardt2017}. 
However, in principle it can be realized directly by means of an external force applied to the atoms using magnetic field or light intensity gradients 
\cite{semeghini2018,masi2021bnsl}, whose slope is varied periodically in time. In this scenario, the potential \eqref{eq:pott} emerges directly in the laboratory reference frame and does not require a transformation to a comoving frame, since the BNSL lattice itself, namely the double well, is static.

By suitably choosing the parameters $F_0$ and $\omega$, the emergent dynamics can be restricted to an effective two-level system. 
In particular, the drive frequency is fixed to $\nu=\omega/2\pi=50$ Hz, with $\nu_{12}\ll \nu\ll \nu_{23}$, so that the modulation is fast compared to the characteristic time scale of the two-level dynamics while avoiding resonances with higher levels. 
The force amplitude is chosen as $F_{0}=1.2\times 10^{-26}$ N, yielding $K=F_{0} a/(h\nu)=\pi/2$, which is the condition for generating rotations about the $y$-axis of the Bloch sphere, as will become clear later in the context of the two-mode Floquet approach. 
It is worth noting that, in the presence of the driving, the Rabi period of the oscillations between the two sub-wells is renormalized to $T'_R\simeq 1280$ ms.

\begin{figure}[t] 
\centerline{\includegraphics[width=\columnwidth]{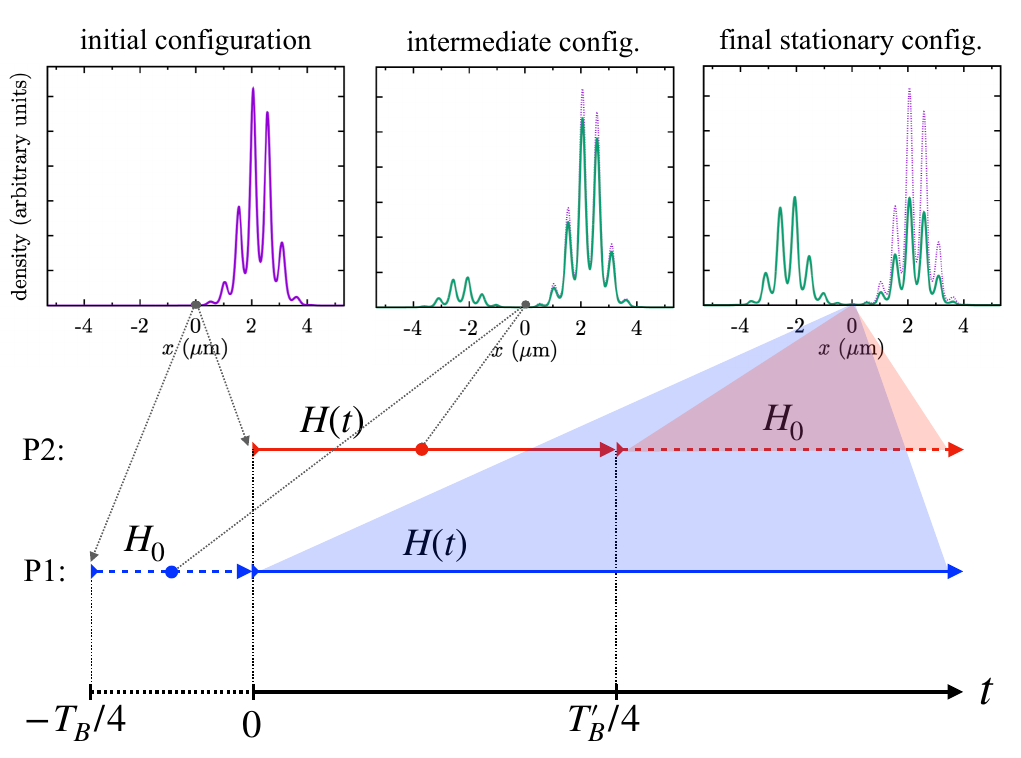}}
\caption{Scheme of the two protocols P1 and P2 employed in this work (see text), indicated by blue and red lines, respectively. The top row displays three snapshots of the density configurations corresponding to the initial state (at different times for the two protocols), an intermediate configuration, and the final stationary configuration with equal population in the two wells.}
\label{fig2}
\end{figure}

In the following, we consider two stop-and-go protocols, each enabling rotations about specific axes on the Bloch sphere, as shown in Fig.~\ref{fig2}:

Protocol 1 (P1): The system is initialized at $t=-T_{B}/4$ in an eigenstate of $\sigma_{z}$, namely $\psi_{L}$ or $\psi_{R}$ (all atoms occupying one of the two sub-wells), and allowed to evolve until $t=0$, when the two wells are equally populated. At this point, the drive is switched on according to \eqref{eq:pott}. 

Protocol 2 (P2): The system is prepared in the same initial state, but at $t=0$ -- when the drive is switched on. Here, the drive is applied only for a finite duration, until $t\simeq T'_R/4$, when the population imbalance vanishes. At that point, the drive is switched off, and the population remains frozen thereafter, with the dynamics governed solely by $H_0$.

\begin{figure}[t] 
\centerline{\includegraphics[width=0.98\columnwidth]{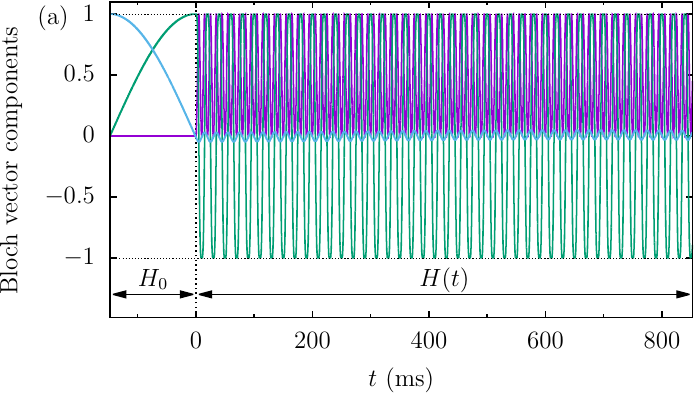}}
\vspace{12pt}
\centerline{\includegraphics[width=0.98\columnwidth]{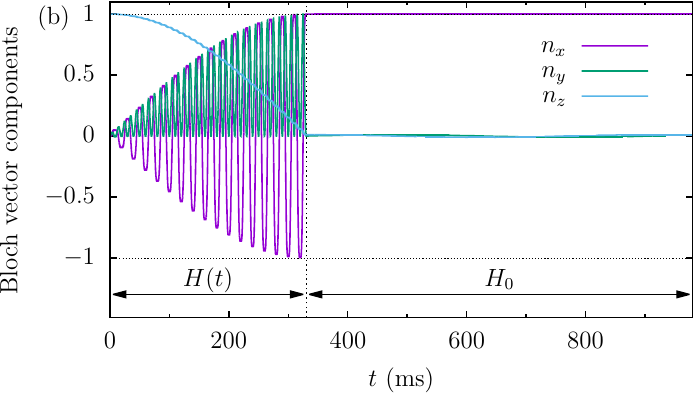}}
\caption{Time evolution of the components of the Bloch unit vector for the two protocols, P1 (a) and P2 (b). See text for details.}
\label{fig:bloch}
\end{figure}

The Bloch sphere representation introduced in the previous section provides a convenient framework to understand how the two protocols work. The components of the Bloch unit vector can be computed from the state $|\psi(t)\rangle$ by standard projections, defining $c_{L,R}(t) = \langle\psi_{L,R}|\psi(t)\rangle$, as follows:
\begin{align}
n_x(t) &= 2 \operatorname{Re}[c_{L}^{*}(t)c_R(t)] \\
n_y(t) &= 2 \operatorname{Im}[c_{L}^{*}(t)c_R(t)] \\
n_z(t) &= |c_L(t)|^2 - |c_R(t)|^2.
\end{align}
The corresponding evolution for the two protocols is shown in Fig.~\ref{fig:bloch}.

P1, in panel (a), begins with a rotation about the $x$-axis, bringing the state to an eigenstate of $\sigma_y$ at $t = 0$. 
At this point, under the effect of the drive, the Bloch vector rotates about the $z$-axis, undergoing rapid oscillations in the $xy$-plane and thereby keeping the populations in the two sub-wells nearly equal.
In P2, panel (b), the Bloch vector progressively tilts away from the $z$-axis toward the $xy$-plane, while undergoing rapid transverse oscillations. At $t = T'_R/4$, the population imbalance vanishes and the Bloch vector aligns along the $x$-axis. For $t > T'_R/4$, the dynamics is governed solely by $H_0 \propto \sigma_x$, so the Bloch vector direction remains unchanged and the populations in the two sub-wells remain frozen.

These two protocols demonstrate that the drive enables controlled rotations around orthogonal axes of the Bloch sphere. A more transparent understanding of the underlying mechanisms is obtained within a simplified framework, as discussed below.

\section{Two-level model}
\label{sec:tlm}

To gain further insight into the behavior of the protocols discussed in the previous section, we introduce a two-level description of the driven system, in analogy with the treatment of the static model in Eq.~\eqref{hamsigmax}. This choice is justified by the fact that the driving regime we consider is characterized by a frequency $\nu$ satisfying $\nu_{12} \ll \nu \ll \nu_{23}$, so that the modulation is fast compared to the intrinsic time scale of the two-level dynamics while avoiding resonances with higher levels \footnote{In the numerical simulations, $|\bm{n}(t)|^2=1$ up to one part in $10^4$, confirming a genuine two-level dynamics.}. In this regime, the full time-dependent Hamiltonian $H(t)$ in Eq.~\eqref{hamtimeperiodic} can be mapped onto the effective two-level form
\begin{equation}
H^{\mathrm{tl}}(t) = -J\sigma_x + A\sin(\omega t)\sigma_z,
\label{hsigmaxezeta}
\end{equation}
where $A = F_0 a / 2$. Within this setting, one can derive an approximate expression for the full time-dependent evolution operator, as described below, by means of a Floquet-based approach that is not restricted to stroboscopic dynamics and accurately captures the fast evolution within each driving period. It also provides a natural framework to clarify the choice of $K=F_{0} a/(h\nu)=\pi/2$. In the following, we first construct the evolution operator within this framework and then use it to analyze the resulting two-level dynamics.

\subsection{Evolution operator}
\label{sec:uevol}

In case of free evolution under the Hamiltonian \eqref{hamsigmax}, the evolution operator is 
\begin{equation}
U_{0}(t,t_{i}) = e^{\displaystyle{i}J\sigma_{x}(t-t_{i})/\hbar} \equiv e^{\displaystyle i\theta(t,t_{i})\sigma_x},
\end{equation}
where $\theta(t,t_{i}) = \Omega_{0}(t-t_{i})$, with $\Omega_{0} \equiv J/\hbar$, and $t_i$ a generic initial time.
It can be written in a $2\times2$ matrix representation as 
\begin{equation}
U_0(t,t_{i}) = \begin{pmatrix}
\cos\theta(t,t_{i}) & i\sin\theta(t,t_{i}) \\
i\sin\theta(t,t_{i}) & \cos\theta(t,t_{i})
\end{pmatrix}.
\label{barevol}
\end{equation}

In the presence of the drive, the evolution operator $U(t,t_{0})$ can be factorized as \cite{bukov2015,holthaus2016}
\begin{equation}
U(t,t_{0}) = P(t,t_{0})e^{-\displaystyle{i} H_F[t_0](t-t_{0})/\hbar},
\label{eq:ufact}
\end{equation}
where $H_F[t_0]$ is the time-independent effective Floquet Hamiltonian, defined through the one-period \textit{stroboscopic evolution operator}, 
\begin{equation}
U(t_0+T,t_0)={\cal{T}}e^{-\displaystyle\frac{i}{\hbar}\int_{t_{0}}^{t_{0}+T}\!\!\!\!\!\! H(t)dt}
\equiv e^{-\displaystyle\frac{i}{\hbar} H_F[t_0]T},
\end{equation}
and $P(t,t_{0})$ is a periodic operator, $P(t,t_{0}+nT) = P(t,t_{0})$, associated with the fast dynamics within a single driving period. 
In the following, since we are interested in the dynamics generated by the drive, we set $t_{0}=0$, corresponding to the initial time of the driving protocol.

This decomposition of the evolution operator is particularly useful because it explicitly separates slow and fast dynamics and provides a convenient starting point for a direct computation via a standard Magnus expansion. We start by considering the following unitary transformation of the system states,
\begin{equation}
 |\tilde\psi(t)\rangle=R^{\dagger}(t)|\psi(t)\rangle,
 \end{equation}
 with
\begin{equation}
 R(t)=e^{\displaystyle{iK\cos(\omega t)\sigma_z/2}},
\end{equation}
and $K={F_0 a}/(\hbar\omega)$, as defined in the previous section. 
The transformed state $ |{\tilde{\psi}}(t)\rangle$ obeys to
 \begin{equation}
  i\hbar\partial_{t} |\tilde\psi(t)\rangle = {\tilde{H}}(t) |\tilde\psi(t)\rangle,
 \end{equation}
 with
 \begin{align}
  \label{htildeuno}
 \tilde{H}(t) &= R^{\dagger}(t)H^{\mathrm{tl}}(t)R(t)-i\hbar R^{\dagger}(t){\dot{R}}(t)
 \\
 &=-J\cos[K\cos(\omega t)]\sigma_x-J\sin[K\cos(\omega t)]\sigma_y. \nonumber
 \end{align}
Then, we consider the Floquet Hamiltonian in this representation,
\begin{equation}
\tilde{H}_F=\frac{i\hbar}{T}\log\left[{\cal{T}}\exp\left(-\displaystyle\frac{i}{\hbar}\int_{0}^{T}\tilde{H}(t)\,dt\right)\right].
\end{equation}
Customarily, this is approximated through a high-frequency Magnus expansion as
\begin{equation}
\tilde{H}_F = \tilde{H}^{(0)}_F + \tilde{H}^{(1)}_F + \dots,
\label{eq:HF}
\end{equation}
which, in our case, is justified by the condition $\nu\gg\nu_{12}$ discussed above.
The leading order is \cite{bukov2015}
\begin{equation}
\tilde{H}^{(0)}_F = \displaystyle{\frac{1}{T}}\int_{0}^{T}{\tilde{H}}(t)dt 
=-J {\cal{J}}_0(K)\sigma_x,
\label{heff}
\end{equation}
owing to 
\begin{equation}
\langle \cos[K\cos(\omega t)]\rangle_{T}={\cal{J}}_0(K),
\end{equation}
and $\langle \sin[K\cos(\omega t)]\rangle_{T}=0$.
Note that this result is independent of the choice of $t_{0}$. This is not generally true at subleading orders, which may depend on $t_{0}$. In particular, for $t_{0}=0$, the next-to-leading-order term $\tilde{H}^{(1)}_F$ vanishes exactly (see Appendix~\ref{sec:appendix}). Therefore, up to corrections of order $(\omega_R/\omega)^2$, the stroboscopic effective Hamiltonian is simply given by the zeroth-order expression in Eq.~\eqref{heff}.

Then, going back to the original state representation, the evolution of $|\psi(t)\rangle$ at times $t=nT$ is generated the Floquet Hamiltonian as
\begin{align}
|\psi(nT)\rangle &= R(nT)|{\tilde{\psi}}(nT)\rangle\nonumber\\
&= R(0)e^{-\displaystyle{i}\tilde{H}^{(0)}_FnT/\hbar}R^{\dagger}(0)|\psi(0)\rangle\nonumber\\
&\equiv e^{-\displaystyle{i}H^{(0)}_{F}nT/\hbar}|\psi(0)\rangle,
\label{evopnT}
\end{align}
where we have defined
\begin{align}
H^{(0)}_{F} &= R(0)H_\mathrm{eff}R^{\dagger}(0)\nonumber\\
&= -J{\mathcal{J}}_0(K)\Big[\sigma_x\cos(K)
+\sigma_y\sin(K)\Big].
\label{floqhamlab}
\end{align}
It is worth noting that $H^{(0)}_{F}$ is the generator of rotations about an arbitrary axis in the equatorial plane of the Bloch sphere, whose orientation is controlled by the parameter $K$. Since we are here interested in rotations about the $y$-axis, we fix $K=\pi/2$, as discussed in the previous section, obtaining
\begin{equation}
H^{(0)}_{F}=-J{\cal{J}}_0(\pi/2)\sigma_y = -J_{\mathrm{eff}}\sigma_y,
\label{eq:H0f}
\end{equation}
with $J_{\mathrm{eff}} = J{\cal{J}}_0(\pi/2)$.

Finally, the entire dynamics can be reconstructed by using the full evolution operator in Eq.~\eqref{eq:ufact}.
In particular, the unitary operator $P(t,0)$ can be expressed in terms of the \textit{kick operator} $K_{F}(t)$ as \cite{bukov2015}
 \begin{equation}
 P(t,0) = e^{-\displaystyle{iK_{F}(t)}}.
 \label{fastop}
 \end{equation}
The kick operator, like the Floquet Hamiltonian, can be obtained from a high-frequency expansion, with each term expressed in terms of the Fourier components of the Hamiltonian $H^{\mathrm{tl}}(t)$ in Eq.~(\ref{hsigmaxezeta}). Consistently with the previous procedure, we retain only the leading non-vanishing order, which reads \footnote{See Eq.~(44) of Ref.~\cite{bukov2015}}
 \begin{equation}
 K^{(1)}_F(t) = \displaystyle{\frac{i}{\hbar\omega}}\sum_{n\neq 0} H^{\mathrm{tl}}_{n}\displaystyle{\frac{1-e^{i\omega n}}{n}}.
 \end{equation}
Since the only non vanishing Fourier components are $H^{\mathrm{tl}}_1=A/(2i)\sigma_z=-H^{\mathrm{tl}}_{-1}$, this gives
\begin{equation}
	K_F(t)\simeq -\frac{K}{2}[\cos(\omega t)-1]\sigma_z \equiv -\Phi(t)\sigma_z,
\end{equation}
with $\Phi(t)\equiv (K/2)\left[\cos(\omega t)-1\right]$, yielding $P(t,0)\simeq R(t)R^{\dagger}(0)$.
The evolution operator then reads
\begin{align}
	U(t,0) &= R(t)R^{\dagger}(0)~e^{\displaystyle-iH^{(0)}_{F}t/\hbar}
	\nonumber\\
	&= e^{\displaystyle{i}\Phi(t)\sigma_z}e^{\displaystyle{i}\alpha(t)\sigma_y},
\end{align}
with $\alpha(t)=\Omega t$ and $\Omega\equiv J_\mathrm{eff}/\hbar$. It can be written explicitly as
\begin{equation}
U(t,0)=
\begin{pmatrix}
\cos\alpha(t)\, e^{\displaystyle i\Phi(t)} & \sin\alpha(t)\, e^{\displaystyle i\Phi(t)}\\
-\sin\alpha(t)\, e^{\displaystyle -i\Phi(t)} & \cos\alpha(t)\, e^{\displaystyle -i\Phi(t)}
\end{pmatrix}.
\label{eq:uevol}
\end{equation}

\subsection{Two-level dynamics}
\label{sec:tldynamics}

The full dynamics shown in Fig.~\ref{fig:bloch} for the two protocols P1 and P2 can be accurately reproduced within the two-level basis using the evolution operators derived above.

\textbf{P1.} In this case, the system is prepared at $t_i = -T_R/4 = -\pi/(4\Omega_0)$ in the state $|+z\rangle = (1,0)$ and evolves freely as
\begin{equation}
|\psi(t\leq 0)\rangle =
\begin{pmatrix}
\cos\theta(t,t_{i})\\
i\sin\theta(t,t_{i})
\end{pmatrix}.
\end{equation}
During this evolution, the Bloch vector components satisfy $n_x(t)\equiv 0$, while
\begin{equation}
\begin{cases}
n_y(t) = \sin2\theta(t,t_{i})\\
n_z(t) = \cos2\theta(t,t_{i}).
\end{cases}
\end{equation}
At $t=0$, this yields the state $|+y\rangle = (1,i)/\sqrt{2}$. At this point, the drive is switched on, and the state is propagated by the evolution operator \eqref{eq:uevol} as
\begin{equation}
|\psi(t\geq 0)\rangle = \frac{e^{\displaystyle{i}\alpha(t)}}{\sqrt{2}}
\begin{pmatrix}
e^{\displaystyle{i}\Phi(t)}\\
i e^{\displaystyle-i\Phi(t)}
\end{pmatrix}.
\end{equation}
This specific form follows from the choice $K=\pi/2$, see Eq.~\eqref{eq:H0f}.
The components of the Bloch vector are therefore
\begin{equation}
\begin{cases}
n_x(t) = \sin 2\Phi(t) \\
n_y(t) = \cos 2\Phi(t) ,
\end{cases}
\end{equation}
with $n_{z}(t) \equiv 0$, which accurately reproduce the behavior shown in Fig.~\ref{fig:bloch}(a).
In particular, since $\Phi(nT)=0$, stroboscopic probing at integer multiples of the driving period shows the Bloch vector aligned along the $y$-axis.

\textbf{P2.} In this protocol, the system is prepared in the state $|+z\rangle$ at $t=0$, and then it is immediately subjected to the drive. Its evolution reads
\begin{equation}
|\psi(t\le t')\rangle=
\begin{pmatrix}
\cos\alpha(t) e^{i\Phi(t)} \\
-\sin\alpha(t) e^{-i\Phi(t)}
\end{pmatrix},
\end{equation}
with Bloch components
\begin{equation}
\begin{cases}
n_x(t)=-\sin 2\alpha(t) \cos 2\Phi(t)\\
n_y(t)=\sin 2\alpha(t) \sin 2\Phi(t)\\
n_z(t)=\cos 2\alpha(t).
\end{cases}
\end{equation}
This result corresponds to the behavior shown in Fig.~\ref{fig:bloch}(b), which displays the full time evolution of the system. 
In particular, if one looks at the evolution at integer multiples of the driving period, the system exhibits a slow rotation around the $y$-axis.
The corresponding Rabi period is $T'_R = T_R/{\cal{J}}_0(\pi/2)$, with ${\cal{J}}_0(\pi/2)=0.472$, yielding $T'_R \simeq 1250$ ms. This is in close agreement with the numerical result $T'_R \simeq 1280$ ms, with a discrepancy of only a few percent \footnote{This discrepancy is most likely due to a combination of higher-order corrections and small deviations of the continuous model from the ideal two-level approximation.}.

The drive is then switched off at $t' = T'_R/4 + \delta$, where $\delta$ is chosen such that $t' = (n+1/2)T$. In the limit $\delta \ll 1$, one has $\alpha \simeq \pi/4$, while the choice $K = \pi/2$ fixes $\Phi = -\pi/2$. This combination yields a state close to $|+x\rangle = (1,1)/\sqrt{2}$ (up to a global phase), which remains aligned along the $x$-axis during the subsequent free evolution,
\begin{equation}
|\psi(t>t')\rangle = -{\frac{i}{\sqrt{2}}}\,e^{-i\theta(t,t')}
\begin{pmatrix}
1\\
1
\end{pmatrix},
\end{equation}
as seen in Fig.~\ref{fig:bloch}(b).

This analysis shows that the two-level description captures the main features of the full dynamics for both protocols P1 and P2, including the relative phase accumulation and the resulting rotations on the Bloch sphere. In particular, the evolution operators derived above provide an accurate representation of the dynamics shown in Fig.~\ref{fig:bloch}.

\section{Conclusions}
\label{sec:concl}

We have investigated the coherent dynamics of a non-interacting condensate in a periodically driven double-well potential and shown that suitable stop-and-go protocols enable controlled rotations about arbitrary axes in the equatorial plane of the Bloch sphere. 
Using a Beat-Note Superlattice realization of a double-well system, we demonstrated through numerical simulations that a spatially linear periodic force can be used to control the population imbalance between the two wells and to prepare specific pseudospin states.
The numerical results are accurately reproduced by a two-level Floquet description.  
Within this framework, the full time-dependent evolution operator has been obtained analytically through a high-frequency Magnus expansion. In particular, for a suitable combination of the force strength, its periodicity, and the lattice spacing (i.e., of the parameter $K$), the effective Floquet Hamiltonian generates rotations about the $y$-axis of the Bloch sphere, while the undriven dynamics correspond to rotations about the $x$-axis.
The proposed protocols provide a clean demonstration of how Floquet engineering can be employed to achieve geometric control of double-well pseudospins.
Such geometric control may find direct application in experimental sequences, such as optimized spin-echo schemes, to suppress decoherence in atom interferometry arising from imperfections in beam-splitting pulses and from noise affecting the energy imbalance between the two wells \cite{petrucciani2026}.

%---------------------------------------------------
\begin{acknowledgments}
%---------------------------------------------------

We acknowledge Álvaro R. Puente-Uriona for valuable discussions and contributions in the initial stage of this work, and Nicola Poli for useful discussions. MM acknowledges support from Grants No. PID2021-126273NB-I00 funded by MCIN/AEI/10.13039/501100011033 and by ``ERDF A way of making Europe'', and from the Basque Government through Grant No. IT1470-22. 

\end{acknowledgments}

\appendix
\section{}
\label{sec:appendix}

The next-to-leading-order contribution to the Floquet Hamiltonian in Eq.~\eqref{eq:HF} is 
\begin{align}
{\tilde{H}}^{(1)}_F[t_0] &= \displaystyle{\frac{1}{2i\hbar T}}\int_{t_0}^{t_0+T}dt_1\int_{t_0}^{t_1}dt_2~\left[{\tilde{H}}(t_1),
{\tilde{H}}(t_2)\right]\nonumber\\
&=\displaystyle{\frac{1}{\hbar\omega}}\sum_{n=1}^{\infty}\displaystyle{\frac{1}{n}}\Big([{\tilde{H}}_n,{\tilde{H}}_{-n}]
-e^{in\omega t_0}[{\tilde{H}}_n,{\tilde{H}}_0]\nonumber\\
&\qquad\qquad\qquad\qquad+e^{-in\omega t_0}[{\tilde{H}}_{-n},{\tilde{H}}_0]\Big),
\label{firstorder}
\end{align}
where the last expression is written in terms of the Fourier components of ${\tilde{H}}(t)$,
\begin{equation}
\tilde{H}(t)=\sum_{n=-\infty}^{\infty}{\tilde{H}}_n~ e^{i n\omega t}.
\end{equation}
Using the Jacobi-Anger expansion,
\begin{equation}
e^{\pm iK\cos(\omega t)}=\sum_{n=-\infty}^{\infty}e^{\pm in\frac{\pi}{2}} {\mathcal {J}}_n(K)e^{in\omega t},
\end{equation}
after some algebra one obtains
\begin{equation}
\tilde {H}_n=- J{\mathcal{J}}_n(K)\left[\cos\left(\frac{n\pi}{2}\right)\sigma_x 
+ \sin\left(\frac{n\pi}{2}\right)\sigma_y\right].
\end{equation}
It is straightforward to verify that for $t_0=0$ the next-to-leading-order term in Eq.~\eqref{firstorder} vanishes.
Using ${\cal{J}}_{-n}(K)=(-1)^n{\cal{J}}_{n}(K)$ together with the expression of ${\tilde{H}}_n$, one finds that for odd $n$, ${\tilde{H}}_n={\tilde{H}}_{-n}\propto\sigma_y$, while for even $n$, ${\tilde{H}}_n={\tilde{H}}_{-n} \propto\sigma_x$. It then follows that for $t_0=0$ the first-order contribution in the Magnus expansion vanishes.

\bibliography{biblio.bib}

\end{document}